\documentstyle[11pt,twoside,paspconf,epsf]{article}

\begin{document}

\title{Searching for z = 6.5 Galaxies with Multislit Windows}
\author{David Crampton}
\affil{DAO/HIA, National Research Council of Canada, Victoria, V8X 4M6 Canada.
Email: David.Crampton@hia.nrc.ca}
\author{Simon Lilly}
\affil{Department of Astronomy. University of Toronto, Toronto, Canada}
\begin{abstract}
A method for searching for emission-line objects in ``windows" between
atmospheric emission lines using multislits is described. A search for
Ly $\alpha$ emitters at z = 6.5 in the 9130\AA\ window using this
technique is being carried out with the multi-object spectrograph on
CFHT. This technique could be extended to similar windows at longer
wavelengths, aided by the (1 + z) factor in observed equivalent
widths.  In the $J$ band there are windows corresponding to Ly $\alpha$
at z = 7.7, 8.7 and 9.3; in the $H$ band, there are two at z = 11.9 and
13.4.  Multislit window observations in these bands coupled with
photometric redshift information offer perhaps the best method of the
detecting extremely high redshift galaxies.
\end{abstract}


\section{Introduction}
Questions of how galaxies and quasars formed in the early Universe and
when and how the intergalactic medium became ionized continue to occupy
a central place in observational cosmology. As quasars and galaxies are
discovered at higher and higher redshifts, the questions posed by these
primordial objects become more pressing, offering the strongest
constraints on cosmogonic theory (e.g., Silk \& Rees 1997, Haehnelt et
al. 1998). The detection of several galaxies with z $>$ 5 have recently
been reported (e.g., Cowie \& Hu 1998; Dey et al. 1998; Hu, Spinrad and
Lanzetta in these proceedings) indicating that searches for objects at
even higher redshift are essential. Most of the highest redshift
galaxies have been detected as a result of their emission lines, either
through targeted narrow band filter surveys (e.g., Cowie \& Hu 1998,
Hu, Cowie \& McMahon 1998), as a result of being lensed by a foreground
cluster (e.g., Franx et al. 1997), or serendipitously in deep, long
slit exposures (e.g., Hu, Cowie \& McMahon 1998, Dey et al. 1998).

Unfortunately, as Ly $\alpha$ moves to higher and higher wavelengths
the atmospheric emission lines imposed on the object spectra become
stronger and stronger and hence harder to remove. Increased fringing
and decreasing detective quantum efficiency in CCD detectors
considerably exacerbates this problem. Hence, despite the fact that
very strong Ly $\alpha$ emission is frequently observed in high
redshift galaxies, such lines are usually only detected in line-free
regions. There are not many such ``windows" that are very wide, but
there is nice one between $\sim$8965\AA\ $-$ 9290\AA\ which is the subject
of this investigation. These wavelengths correspond to redshifted 
Ly~$\alpha$ at 6.38 $<$ z $<$ 6.64, but also to redshifted [OII] at z
$\sim$ 1.45, [OIII] at z $\sim$ 0.82, and H $\alpha$ at z $\sim$ 0.39.
Fortunately, most of the latter typically have substantially lower
equivalent widths than Ly $\alpha$ and, according to Cowie \& Hu (1998)  most
observed emission lines with W~$>>$ 100\AA\ are Ly~$\alpha$. Fortuitously,
atmospheric water vapour absorption is low in this
window.

Apparently, the density of high redshift Ly $\alpha$  emitters is very
high. Hu et al. (1998) estimate that there are 15,000 per square
degree per unit z with V $<$ 25.25, based on narrow band emission line
surveys at z = 3.4 and z = 4.6. These galaxies are very faint, of
course, but, in general, they are very compact and have narrow emission
lines, both of which aid in their detectability by spectroscopic
means.

\section{Observations}
\subsection{Strategy}

To search for strong Ly $\alpha$ emitters, we decided to use a grism
giving 10.5\AA\ per pixel on the CFHT multi-object spectrograph in
multi-slit mode in conjunction with a custom filter 300\AA\ wide
centered at 9130\AA. To survey a reasonably large area yet maintain a
low sky brightness, 2\arcsec\ wide slits (corresponding to five
0\farcs4 pixels) were used (it should be noted, however, that the
small size of very high redshift galaxies means that the spectra are
essentially slitless and, with typical seeing at CFHT of $<$ 1\arcsec,
the effective resolution is about 25\AA). The 300\AA\ long spectra
occupy about 29 pixels at our chosen resolution. However, the wings of
the filter transmission curve are not square and so an additional 10
pixels were allocated to avoid any overlapping of spectra.  Since the
total height covered by the slit mask corresponds to 1150 pixels, 30
spectra (from 30 long slits) can be fitted into the available area. The
precise slit separation was chosen so that the zero orders from slits
at the top of the mask would fall within the 10 pixel ``inter-spectral''
gaps which were allowed for the filter wings. Two small ``bridges" were
left in the slits to ensure the integrity of the mask after all the
slits were cut. An image of the mask is shown in Figure 1a.

\begin{figure}
\plottwo{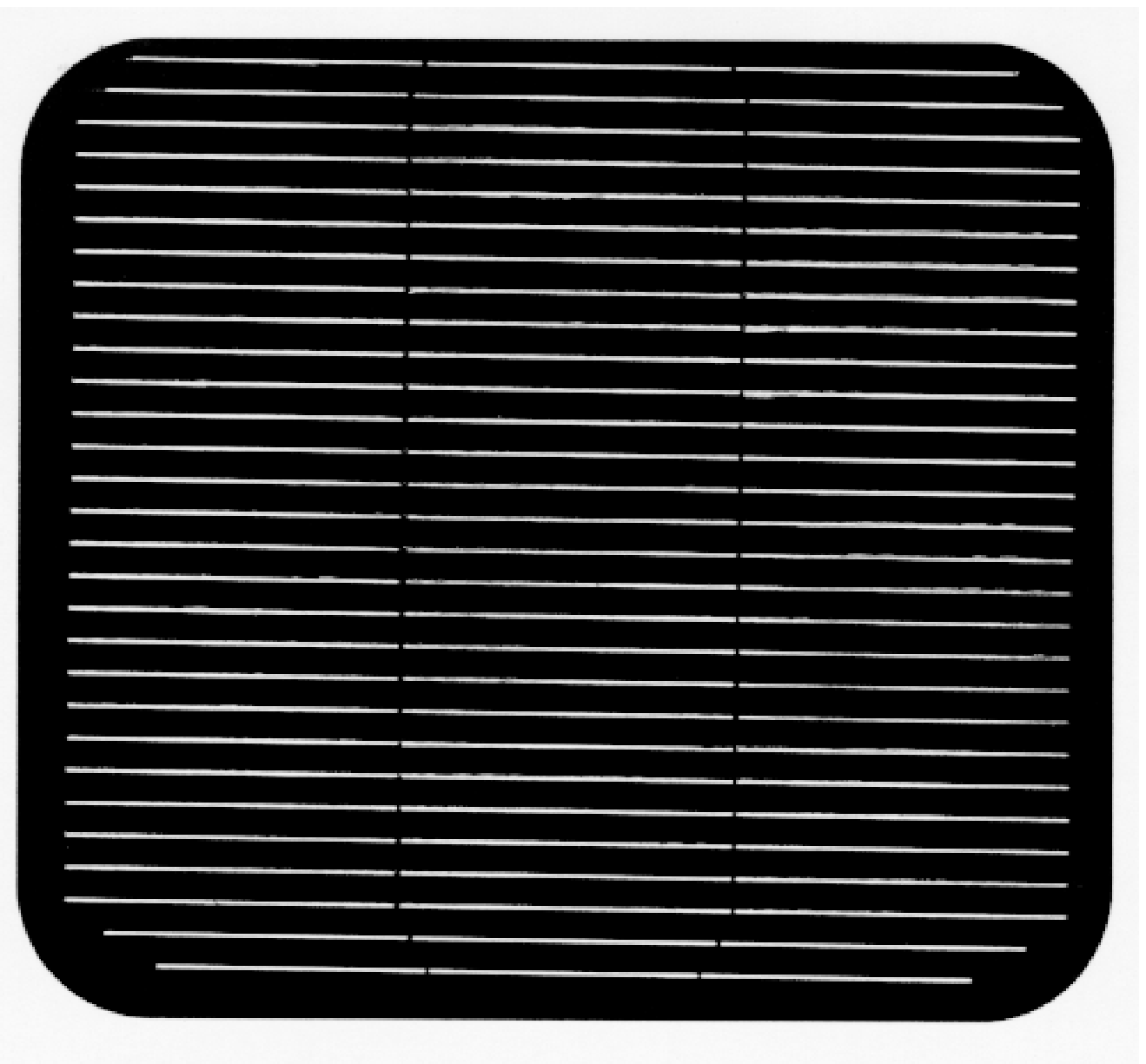}{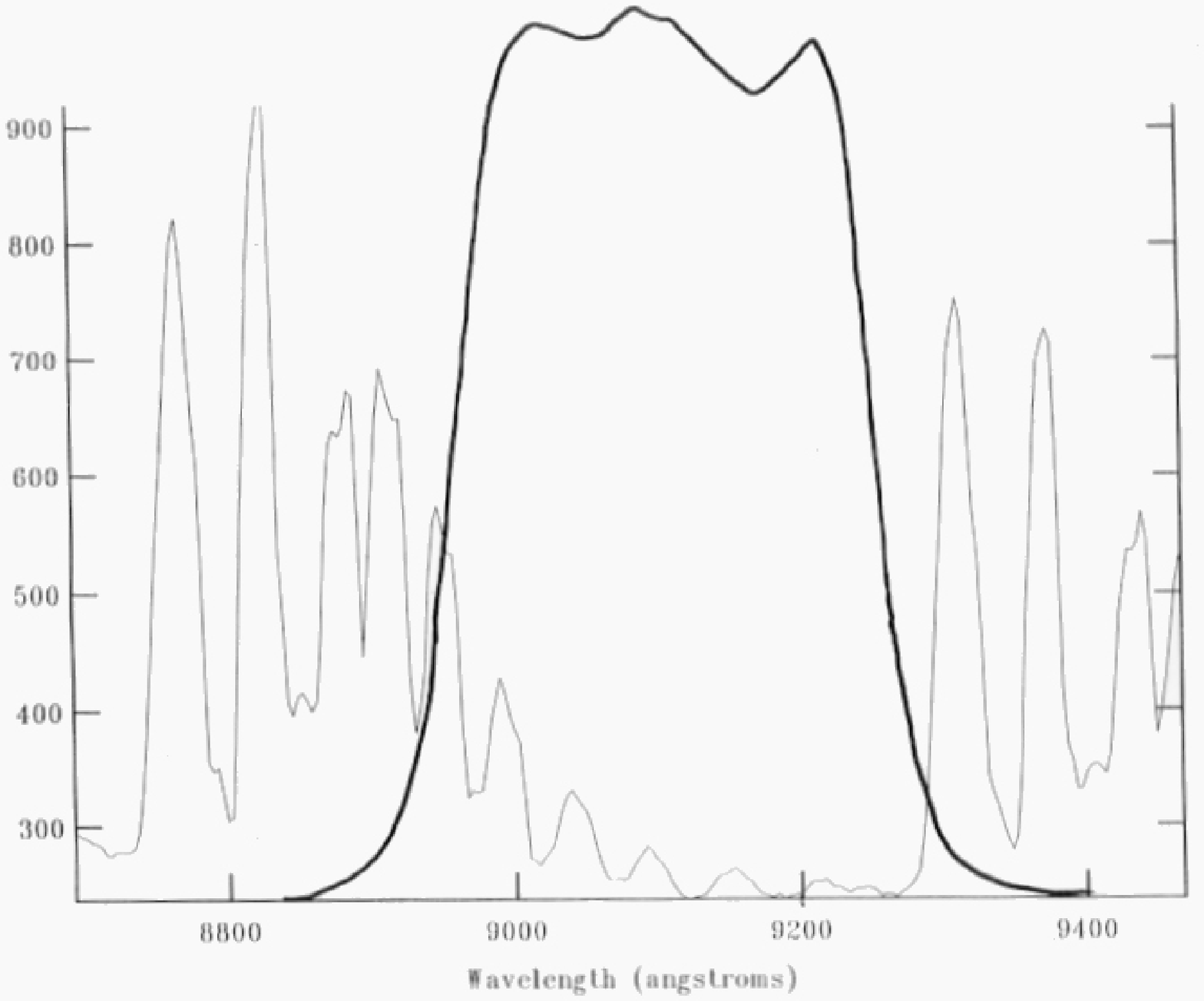}
\caption{a) (Left) A picture of the mask showing the multislits. Two
small bridges were used in each slit to maintain the integrity of the mask. (b)
The transmission curve of the custom filter is shown superimposed
on a spectrum of the night sky near 9130\AA. Note the increased sky background
in the filter passband at shorter wavelengths} 
\end{figure}

The total area covered by the slits is 2\arcsec\ * 9\arcmin\ long * 30
slits or 9 square arcmin. This is only about 1/8 the total area of the
focal plane which would be available if instead a narrow band filter
(say 50\AA\ wide) were used for such a survey.  However, with the
multislit technique a six-fold multiplex gain is realized since
300\AA\ are searched simultaneously and, as mentioned above, the
resolution is better, about 25\AA\ for seeing-limited sources.  The
resulting higher sensitivity, albeit over a smaller area, is an
attractive trade given the expected high density of sources. Based on
the density estimated by Hu et al. (1998), approximately 10 Ly $\alpha$
emitters per field might be expected with V $<$ 25.25 in our redshift interval,
although their estimate was based on lower redshift objects.

\subsection{Observations and Reduction Procedure}

Observations of two CFRS fields were carried out with the CFHT in 1998
August.  The bandpass of the 300\AA\ filter was not optimal, being
centered somewhat blueward of the desired wavelength. Figure 1b shows
the transmission of the filter as a function of wavelength,
superimposed on a spectrum of the night sky. The filter has an average
peak transmission of about 90\%, the bandpass is relatively rectangular
in shape, and it has good rejection outside of the transmissive
bandpass. However, as the figure demonstrates, it ideally should have
been centered redward by $\sim$40\AA.  The filter characteristics were
within specifications, but these had been somewhat relaxed to ensure
quick delivery.

Since the precise filter characteristics and precise offsets of the
zero order spectra relative to the first order spectra (in order to
place the zero orders within inter-spectral strips) were not known
in advance of the observations, the slit mask had to be fabricated
immediately prior to the target observations. In order to cut thirty
2\arcsec\ $\times$ 9\arcmin\ slits at normal cutting speeds at CFHT, four hours
are required.  Since this was impractical, the cutting speed was
doubled with the result that the edges of the slits were much rougher
than usual or optimal. Consequentally, the slit mask was far from ideal.
However, our observational technique minimized the associated
problems: the objects were dithered along the slits from one
observation to the next, resulting in excellent flat-field correction.


Series of one hour exposures were obtained of the CFRS1415+52 and
2215+00 fields during three nights for a total of 4h and 16h
respectively. A part of one such exposure is shown in Figure 2a. Short,
300\AA\ long, spectra of a few brighter objects can be seen
superimposed on the background which increases significantally towards
the blue, as expected from the sky background illustrated in Figure
1b.  During observations, a few bright objects were recentered
vertically in the slits before each subsequent exposure and the fields
were offset horizontally by $\sim$2\arcsec\ between exposures.

\begin{figure}
\plottwo{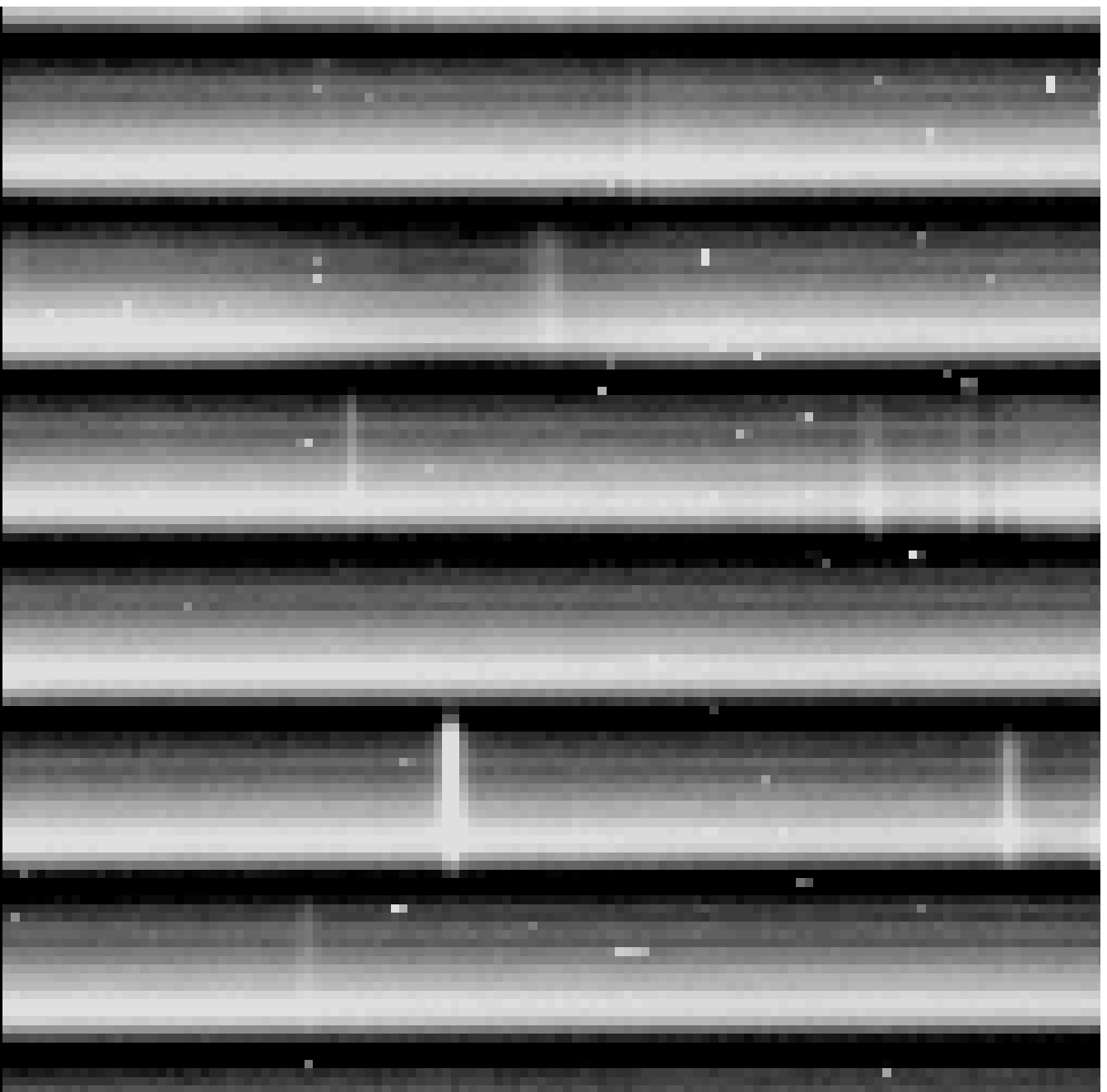}{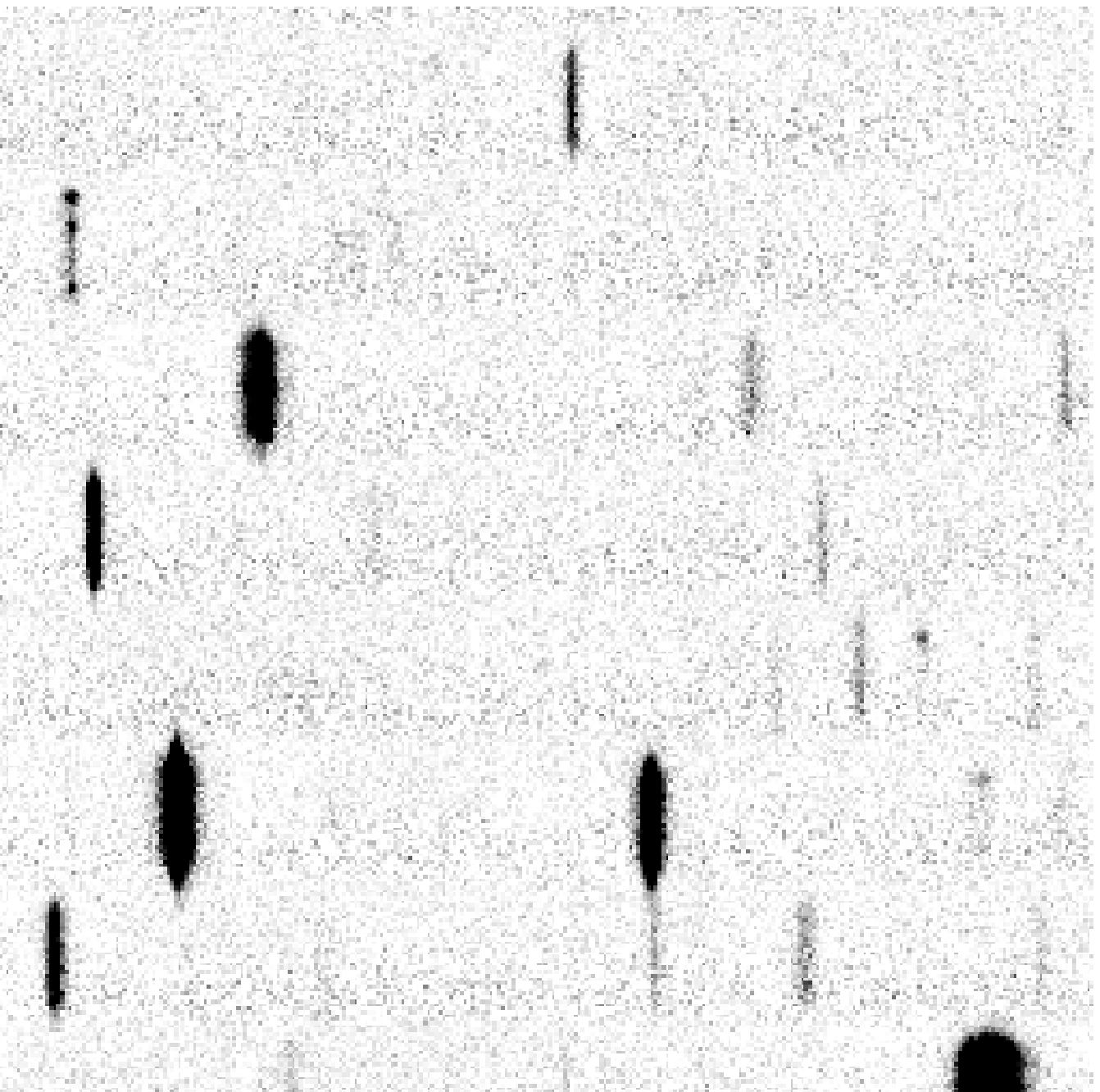}
\caption{a) (left) A portion of a raw spectral image. The background
sky lines are much brighter in the bluer part of the ``window"
(wavelength increases upwards in each spectral strip). Some of the
brighter spectra and cosmic rays are visible superimposed on the
background. Note that the inter-spectral strips are quite dark. (b) A
portion of an image formed by shifting and coadding a series of
flat-fielded and sky background-subtracted spectra. This figure
demonstrates the excellent flatfield correction obtained and shows three
relatively bright emission-line spectra as well as continuum spectra.}
\end{figure}

Individual exposures were first divided by a superflat which was formed
from all the dithered data. The sky background was then subtracted
along the slits in a more or less conventional way with IRAF. These
background-subtracted images were shifted and combined using a badpixel
mask to reject regions of the slit bridges and bad CCD columns. A
portion of the resulting, co-added, spectral image is shown in Figure
2b. As mentioned above, despite the relatively poor slits and the
omnipresent CCD fringing at such wavelengths, the images flat-fielded
extremely well as a consequence of our dithering procedure. Emission
line spectra are clearly visible on these images in addition to the
more common short continuum spectra. 

In principle, recognition software such as Sextractor(Bertin \& Arnouts
1996) could now be used to detect the emission-line objects in an
unbiased way to a given flux threshold.  However, the presence of more
or less point-like zero-order images in the bottom third of the field
(from bright objects in the top third) introduces a considerable number
of spurious images. These can, of course, be identified and removed
{\it post facto}. However, in addition, emission lines superimposed on
continua may also not be recognizable by standard detection software
(although these are unlikely to be the Ly $\alpha$ emitters).  Hence,
two further steps were taken to alleviate these problems: the (known)
inter-spectral strips where the zero orders lay were replaced with
strips of uniform background which have appropriate noise
characteristics, and background subtraction was performed {\it along}
the spectral slices in order to subtract the continua. This process
works well for all spectra but those of the brightest objects (which
are not of interest for this investigation) for which residuals
frequently remain at the ends of the spectra. A portion of a
continuum-subtracted image is shown in Figure 3a (this image was
gaussian smoothed with the resolution expected for point sources to
further enhance such features). Several faint sources in addition to the
three obvious ones (and noise spikes!) are visible. The spurious
residuals from bright stars are disconcerting but easily rejected. With
such images, conventional software can be used to detect the emission
lines in an unbiased manner.

\begin{figure}
\plottwo{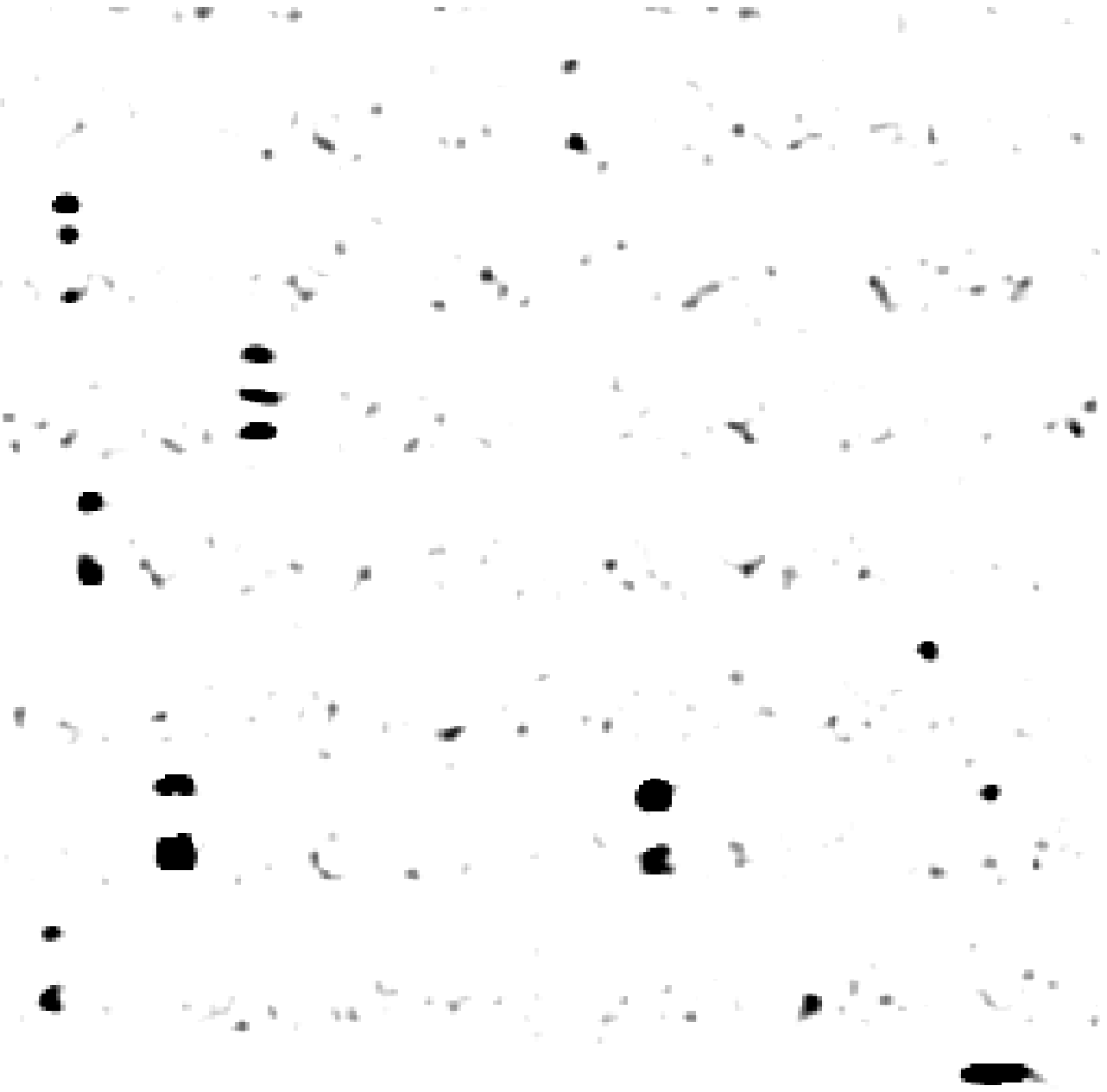}{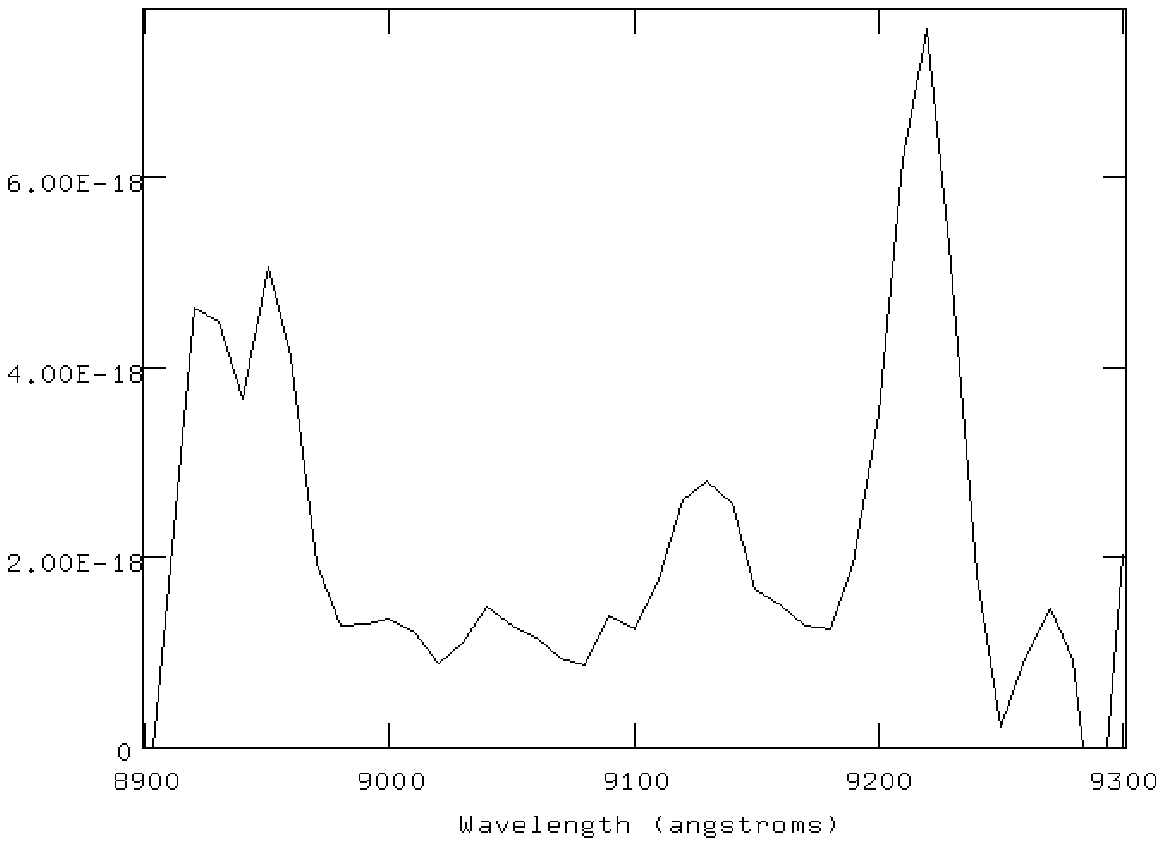}
\caption{a) (left) The same image as in Figure 2b but with the continua
subtracted to improve detection of emission lines superimposed on
continuua. This image was gaussian smoothed to enhance the visibility
of sources for display purposes. Residuals from the brighter continuum
sources are evident but easily rejected. Some potential fainter sources
are visible in this figure in addition to the three obvious ones. (b)
Wavelength and flux calibrated spectrum of the three-emission-line
object at the upper left of Figure 2b and 3a. H$\beta$ (plus noise on the 
violet side) and the [OIII] lines are visible.}

\end{figure}

\section{Emission-line Objects}

Examples of obvious emission-line objects are shown in Figure 3a. Since
this technique produces actual spectra, albeit over a small wavelength
range, actual wavelengths, equivalent widths, and fluxes can also be
extracted and reduced by conventional means. A fluxed spectrum of the
emission-line galaxy shown at the upper left in Figures 2b \& 3a is
reproduced in Figure 3b. This object is clearly a galaxy at z = 0.86
since it shows H $\beta$ and the [OIII] lines. This galaxy is
CFRS22.0184 with I$_{AB}$ $\sim$ 22.4. The two most obvious single
emission line galaxies in Figures 2b \& 3a (lower right) have I$_{AB}$
$\sim$25 and have W $>>$ 100\AA, but it is not obvious without further
evidence (e.g., a photometric redshift) whether they are at z $\sim$
1.4 or $\sim$6.5. Approximately 50 such objects, most of which are much
fainter, are present on our deep exposure of the 22h field.  Analyses
of their colors and further spectroscopic observations are currently
underway to establish their redshifts.

\section{Summary}

A combination of multislits and intermediate band filters appears to be
an excellent method for the detection of faint emission-line objects.
This method has been used to explore the window between the strong
atmospheric emission lines near 9130\AA. Several dozen emission-line objects
have been detected, some of which have equivalent widths
$>>$100\AA\ and hence may be galaxies with redshifted Ly
$\alpha$ emission at z $\sim$ 6.5. Obviously, this technique could be
extended to similar ``windows" at longer wavelengths, aided by the (1 +
z) factor in observed equivalent widths. In the $J$ band there are
windows corresponding to Ly $\alpha$ at z $\sim$ 7.7, 8.7 and 9.3. There
appear to be only two in the $H$ band, corresponding to z $\sim$ 11.9 and
13.4.  Multislit observations in these windows when coupled with
photometric redshift information offer perhaps the best method of the
detection of extremely high redshift galaxies.

Although our initial observations using this technique were successful,
they could be improved upon in several ways. The detective quantum
efficiency of the CCD was reasonable, being about 50\% at this wavelength, but it could be
higher. Similarly, the peak efficiency of the grating occurs at 7500\AA\ 
and so a grating with a more appropriate blaze could be obtained.
Finally, as mentioned above, neither the
filter nor the actual slit mask used were optimal. Taken all together,
a factor of at least two improvement in depth could easily be
achieved.

\acknowledgments

We thank CFHT Director Pierre Couturier for continued moral support and for
agreeing to purchase the requisite filter on short notice.

\end{document}